\documentstyle[12pt]{article}

\let\ni=\noindent

\begin{document}

\baselineskip 0.75cm
 
\pagestyle {empty}

\renewcommand{\thefootnote}{\fnsymbol{footnote}}

\newcommand{\CKM}{Cabibbo---Kobayashi---Maskawa }

~~~

\vspace{1.0cm}

{\large\centerline{\bf Texture dynamics including potential sterile neutrino%
\footnote{Work supported in part by the Polish KBN--Grant 2--B302--143--06.}}}

\vspace{0.8cm}

{\centerline {\sc Wojciech Kr\'{o}likowski}}

\vspace{0.6cm}

{\centerline {\it Institute of Theoretical Physics, Warsaw University}}

{\centerline {\it Ho\.{z}a 69,~~PL--00--681 Warszawa, ~Poland}}

\vspace{0.9cm}

{\centerline {\bf Abstract}}

\vspace{0.3cm}

 A unified form of mass matrix proposed previously for all fundamental fermions
is extended to include the sterile neutrino $\nu_s $, presumed to mix mainly
with $\nu_e $, while $\nu_\tau $ mixes strongly with $\nu_\mu $ (and both $
\nu_\mu $ and $\nu_\tau $ weakly with $\nu_e $ and $\nu_s $). It turns out that
the former mixing can be responsible for oscillations of solar neutrinos, while
the latter for those of atmospheric neutrinos. It is mentioned in a footnote 
that $\nu_s $ is one of two sterile neutrinos, possible from a viewpoint of an 
algebraic construction. The charged--lepton factor $ U^{(e)}$ in the leptonic 
four--dimensional \CKM matrix $ V = U^{(\nu)\,\dagger}U^{(e)}$ is determined 
through the small deviation of our previous prediction $ m_\tau = 1776.80$~MeV 
(valid at the level of $ U^{(e)} = {\bf 1}$) from the experimental value $ 
m_\tau = 1777.00^{+0.30}_{-0.27}$~MeV. Then, it gives corrections to the 
four--dimensional mixing of $\nu_e $, $\nu_\mu $, $\nu_\tau $, $\nu_s$ leading,
in particular, to small $\nu_e \rightarrow \nu_\mu $, $\nu_e \rightarrow 
\nu_\tau $ and $\nu_\mu \rightarrow \nu_s $ oscillations (absent at the level 
of $ U^{(e)} = {\bf 1}$). However, the LSND estimates of $\bar{\nu}_\mu 
\rightarrow \bar{\nu}_e $ and $\nu_\mu \rightarrow \nu_e $ oscillations are 
too large to be explained by these corrections, though in the case of $\nu_\mu
\rightarrow \nu_e $ they seem to be at the edge of such possible explanation.

\vspace{0.3cm} 

\ni PACS numbers: 12.15.Ff , 12.90.+b 
 
\vspace{2.0cm} 

\ni March 1998

\vfill\eject

\pagestyle {plain}

\setcounter{page}{1}

~~~

\vspace{0.3cm}

\ni {\bf 1. Introduction}

\vspace{0.3cm}

 Recently, in the course of extended studies on the "texture" of charged 
leptons $ e^-\,,\,\mu^-\,$, $\tau^- $, neutrinos $\nu_e $, $\nu_\mu $, $
\nu_\tau $, up quarks $ u\,,\,c\,,\,t $  and down quarks $ d\,,\,s\,,\,b $, we 
came to a proposal of unified algebraic structure of their mass matrices $
\left({M}^{(f)}_{ij}\right)$ $(f = \nu\,,\,e\,,\,u\,,\,d)$ in the three--dimen%
sional family space $(i,j = 1,2,3)$ [1]. The proposed structure followed from 
two sources. First of all, from {\it (i)} an idea about the origin of three 
fundamental--fermion families as a consequence of some generalized Dirac--type 
equations (interacting with the Standard--Model gauge bosons) whose {\it a 
priori} infinite series is, due to an intrinsic Pauli principle, reduced (in 
the case of fermions) to three such equations{\footnote{These imply the 
existence of three Dirac bispinors $\psi_{\alpha}^{(1)}$, $\psi_\alpha^{(2)} = 
(1/4)(C^{-1}\gamma^5)_{\beta_1\,\beta_2}\psi_{\alpha\,\beta_1\,\beta_2}$ and $
\psi_\alpha^{(3)} = (1/24)\epsilon_{\beta_1\,\beta_2\,\beta_3\,\beta_4}\psi_{
\alpha\,\beta_1\,\beta_2\,\beta_3\,\beta_4}$, where $\alpha = 1,2,3,4 $ and $
\beta_i = 1,2,3,4 $ $(i = 1,2,\ldots,N-1 $ with $ N-1 = 0,2,4 $) are Dirac 
bispinor indices ($\alpha $ is correlated with the presence of a whole set of 
Standard--Model charges suppressed in our notation, while $\beta_1,\ldots,
\beta_{N-1}$ are fully antisymmetric). We interpret these bispinors as 
fundamental fermions from three families, corresponding to a given Standard--%
Model signature, {\it i.e.,} as $\nu_e $, $\nu_\mu $, $\nu_\tau $ or $ e^-\,,
\,\mu^-\,,\,\tau^- $ or $ u\,,\,c\,,\,t $ or $ d\,,\,s\,,\,b $.}} [2]. And 
further, from {\it (ii)} an ansatz for the fermion mass matrix expressed in 
terms of the suggested family characteristics [2].

 This proposal in the case of leptons reads

\vspace{-0.2cm}

% rownanie 1
\begin{equation}
\left({M}^{(f)}_{ij}\right) = \frac{1}{29} \left(\begin{array}{ccc} 
\mu^{(f)}\varepsilon^{(f)\,2} & 2\alpha^{(f)} e^{i\varphi^{(f)}} & 0 \\ & & 
\\ 2\alpha^{(f)} e^{-i\varphi^{(f)}} & 4\mu^{(f)}(80 + \varepsilon^{(f)\,2})/9 
& 8\sqrt{3}(\alpha^{(f)} - \beta^{(f)}) e^{i\varphi^{(f)}} \\ & & 
\\ 0 & 8\sqrt{3}(\alpha^{(f)} - \beta^{(f)}) e^{-i\varphi^{(f)}} & 24\mu^{(f)}
(624 + \varepsilon^{(f)\,2})/25 \end{array}\right)
\end{equation}

%\vspace{-0.15cm}

\ni with $f = \nu\,,\,e$, while $\mu^{(f)}\,,\,\varepsilon^{(f)\,2}\,,\,
\alpha^{(f)}\,,\,\beta^{(f)}$ and $\varphi^{(f)}$ denote real constants to be 
determined from the present and future experimental data for lepton masses and 
mixing parameters ($\mu^{(f)}\,,\,\alpha^{(f)}$ and $\beta^{(f)}$ are mass--%
dimensional).

 On the base of a numerical experience we assumed that

\vspace{-0.2cm}

% rownanie 2
\begin{equation}
\varepsilon^{(\nu)\,2} = 0\,,\,\alpha^{(\nu)} = 0\,,\,\beta^{(e)}= 0\,,
\end{equation}

%\vspace{-0.15cm}

\ni what leads to $M^{(\nu)}_{11} = 0 $ and $M^{(\nu)}_{12} = 0 = M^{(\nu)}_{21
}$. Then, we determined the parameters $\mu^{(e)}\,$, $\varepsilon^{(e)\,2}$ 
and $\alpha^{(e)}$ from the experimental values of $ m_e\,,\;m_\mu$ and $m_\tau
$, while $\mu^{(\nu)}$ and $\beta^{(\nu)}$ --- from the possible atmospheric--%
neutrino oscillations as seen by Super--Kamiokande. The phases $\varphi^{
(\nu)}$ and $\varphi^{(e)}$ remained undetermined. 

 In fact, in the case of charged leptons, assuming that the off--diagonal 
elements of the mass matrix $ \left({M}^{(e)}_{ij}\right)$ can be treated as a 
small perturbation of its diagonal terms, we calculate in the lowest 
(quadratic) perturbative order [1]

\vspace{-0.3cm}

%rownanie 3
\begin{eqnarray}
m_\tau & = & \frac{6}{125} \left(351 m_\mu - 136 m_e \right) \nonumber \\
& + & \frac{216 \mu^{(e)}}{3625} \left( \frac{111550}{31696 + 29
\varepsilon^{(e)\,2}} - \frac{487}{320 - 5\varepsilon^{(e)\,2}}\right)\,
\left(\frac{\alpha^{(e)}}{\mu^{(e)}}\right)^2 \;,\nonumber \\
\varepsilon^{(e)\,2} & = & \frac{320 m_e}{9 m_\mu - 4 m_e} + O\left[\left(
\frac{\alpha^{(e)}}{\mu^{(e)}}\right)^2 \right] \;, \nonumber \\
\mu^{(e)} & = & \frac{29}{320} \left(9 m_\mu - 4 m_e \right)   + O\left[\left(
\frac{\alpha^{(e)}}{\mu^{(e)}}\right)^2\right]\mu^{(e)} \;.
\end{eqnarray}

\vspace{-0.1cm}

\ni When the experimental $ m_e $ and $ m_\mu $ [3] are used as inputs, Eqs. 
(3) give [1]

\vspace{-0.3cm}

%rownanie 4
\begin{eqnarray}
m_\tau & = & \left[ 1776.80 + 10.2112 \left(\frac{\alpha^{(e)}}{\mu^{(e)}}
\right)^2\,\right]\;{\rm MeV} \;, \nonumber \\
\varepsilon^{(e)\,2} & = & 0.172329 + O\left[\left(\frac{\alpha^{(e)}}{\mu^{(e)
}}\right)^2 \right] \;,\nonumber \\
\mu^{(e)} & = & 85.9924\;{\rm MeV} + O\left[\left(\frac{\alpha^{(e)}}{\mu^{(e)}
}\right)^2\,\right]\,\mu^{(e)}\;. 
\end{eqnarray}

\vspace{-0.1cm}

\ni We can see that the predicted value of $ m_\tau $ agrees very well with its
experimental figure $ m_\tau^{\rm exp} = 1777.00^{+0.30}_{-0.27}$~MeV [3], 
even in the zero--order perturbative calculation. In order to estimate $\left( 
\alpha^{(e)}/\mu^{(e)}\right)^2 $, we can take this experimental figure as 
another input. Then,

\vspace{-0.3cm}

%rownanie 5
\begin{equation}
\left(\frac{\alpha^{(e)}}{\mu^{(e)}}\right)^2 = 0.020^{+0.029}_{-0.020} \;,
\end{equation}

\vspace{-0.1cm}

\ni so it is consistent with zero.

 For the unitary matrix $\left({U}^{(e)}_{ij}\right)$, diagonalizing the mass 
matrix $\left({M}^{(e)}_{ij}\right)$ according to the relation $ U^{(e)\,
\dagger}\,M^{(e)}\,U^{(e)} = {\rm diag}(m_e\,,\,m_\mu\,,\,m_\tau)$, we
obtain in the lowest (linear) perturbative order in $\alpha^{(e)}/\mu^{(e)}$

\vspace{-0.3cm}

%rownanie 6
\begin{equation}
\left( U^{(e)}_{ij}\right) = \frac{1}{29}\left(\begin{array}{ccc} 
29 & 2\frac{\alpha^{(e)}}{m_\mu}e^{i\varphi^{(e)}} & 0 \\ - 2\frac{\alpha^{(e)}
}{m_\mu} e^{-i\varphi^{(e)}} & 29 & 8\sqrt{3}\frac{\alpha^{(e)}}{m_\tau}
e^{i\varphi^{(e)}} \\ 0 & -8\sqrt{3}\frac{\alpha^{(e)}}{m_\tau}e^{-i
\varphi^{(e)}} & 29 \end{array} \right) \;,
\end{equation}

\vspace{-0.1cm}

\ni where the small $\varepsilon^{(\nu)\,2}$ is neglected.

The case of neutrinos is discussed throughout the next Sections. In particular,
we will determine the parameters $\mu^{(\nu)}$ and $\beta^{(\nu)}$ in Section
4 from Super-Kamiokande results.

 We shall not discuss any {\it a priori} motivation for the proposal (1), 
considering it simply as a detailed conjecture (the interested Reader may look 
for its roots in Refs. [2]). Instead, we allow in this paper for the existence 
of a sterile neutrino $\nu_s $ (blind to the Standard--Model interactions), 
extending the neutrino mass matrix $\left({M}^{(\nu)}_{ij}\right)$ ($ i,j = 1,
2,3 $), as given in Eq. (1), to a $ 4\times 4 $ Hermitian matrix $\left( M^{(
\nu)}_{IJ}\right)$ ($I,J = 1,2,3,4$), where  $ M^{(\nu)}_{IJ} = M^{(\nu)\,*}_{
JI}${\footnote{Roots for
this sterile neutrino may be sought again in the generalized Dirac equations 
(this time, without Standard--Model interactions) whose {\it a priori} infinite
number is, due to the intrinsic Pauli principle, reduced (in the case of 
fermions) to two such equations for the Dirac bispinors $\psi_{\beta}^{(1)}$ 
and $\psi_\beta^{(2)} = (1/6)(\gamma^5\,C)_{\beta\,\beta_4}\epsilon_{\beta_4\,
\beta_1\,\beta_2\,\beta_3}\psi_{\beta_1\,\beta_2\,\beta_3}$ ($\beta_1\,,\;
\beta_2\,,\;\beta_3 $ are fully antisymmetric). In the present paper, $\psi_{
\beta}^{(1)}$ is interpreted as a sterile neutrino of the Dirac type denoted by
$\nu_s $. The existence of $\psi_{\beta}^{(2)}$ as another sterile neutrino of 
the Dirac type is not discussed here (such second potential sterile neutrino 
might mix mainly with $\nu_\mu $, producing an extra disappearance mode of $
\nu_\mu $, slightly correcting the effect of its dominating mode $\nu_\mu 
\rightarrow \nu_\tau $; for some comments {\it cf.} Section 6).}}. To this end,
we supplement the former matrix by seven {\it a priori} unknown elements $ 
M^{(\nu)}_{i4}$, $ M^{(\nu)}_{4j}$ and $ M^{(\nu)}_{44}$. We will assume by 
analogy with $ M^{(\nu)}_{12} = 0 = M^{(\nu)}_{21}$  and $ M^{(\nu)}_{13} = 0 
= M^{(\nu)}_{31}$ that 

\vspace{-0.3cm}

%rownanie 7
\begin{equation}
M^{(\nu)}_{24} = 0 = M^{(\nu)}_{42}\;,\; M^{(\nu)}_{34} = 0 = M^{(\nu)}_{43}\;,
\end{equation}

\vspace{-0.1cm}

\ni but allow for nonzero $ M^{(\nu)}_{14}\!$ and $ M^{(\nu)}_{41}\!$ (as well 
as $ M^{(\nu)}_{11}$ and $ M^{(\nu)}_{44}$) in analogy with nonzero $ M^{(\nu)}
_{23}$ and $M^{(\nu)}_{32}$ (as well as $ M^{(\nu)}_{22}\!$ and $ M^{(\nu)}_{33
}\!$). Note that here $ M^{(\nu)}_{11}\! =\! 0\! $ under our par\-ticular 
assumption (2). If there is no sterile neutrino $\nu_s $, then also $ M^{(\nu)
}_{14}\! = \!0\! = \!M^{(\nu)}_{41}$ and $ M^{(\nu)}_{44}\! = \!0 $, and we 
return to the $ 3 \times 3 $ mass matrix $\left({M}^{(\nu)}_{ij}\right)$ ($i,j 
= 1,2,3 $) discussed in Ref.~[1].

\vspace{0.3cm}

\ni {\bf 2. Neutrino mass states}

\vspace{0.3cm}

 The eigenvalues of the mass matrix $\left({M}^{(\nu)}_{IJ}\right)$ are the 
masses of four neutrino mass states $\nu_1\,,\,\nu_2\,,\,\nu_3\,,\,\nu_4 $. 
They are given as follows

\vspace{-0.2cm}

%rownanie 8
\begin{eqnarray}
m_{\nu_1,\,\nu_4} & = & \frac{M^{(\nu)}_{11}+M^{(\nu)}_{44}}{2} \mp \sqrt{
\left(\frac{M^{(\nu)}_{11}-M^{(\nu)}_{44}}{2}\right)^2 + |M^{(\nu)}_{14}|^2}
\;, \nonumber \\ %& & \nonumber \\
m_{\nu_2,\,\nu_3} & = & \frac{M^{(\nu)}_{22}+M^{(\nu)}_{33}}{2} \mp \sqrt{
\left(\frac{M^{(\nu)}_{22}-M^{(\nu)}_{33}}{2}\right)^2 + |M^{(\nu)}_{23}|^2}
\;.
\end{eqnarray}

\vspace{-0.1cm}

The corresponding unitary matrix $\left({U}^{(\nu)}_{IJ}\right)$, diagonalizing
the mass matrix $\left(M^{(\nu)}_{IJ}\right)$ according to the equality $\,U^{
(\nu)\,\dagger}\, M^{(\nu)}\,U^{(\nu)} = {\rm diag}(m_{\nu_1}\,,\,m_{\nu_2}\,,
\,m_{\nu_3}\,,\,m_{\nu_4})$~, takes the form

%\vfill\eject

\vspace{-0.2cm}

%rownanie 9
\begin{equation}
\left({U}^{(\nu)}_{IJ}\right) = \left(\begin{array}{cccc}
\frac{1}{\sqrt{1+Y^2}} & 0 & 0 & -\frac{Y}{\sqrt{1+Y^2}} e^{i\varphi^{(\nu)}} 
\\ 0 & \frac{1}{\sqrt{1+X^2}} & -\frac{X}{\sqrt{1+X^2}} e^{i\varphi^{(\nu)}} &
0 \\ 0 & \frac{X}{\sqrt{1+X^2}}e^{-i\varphi^{(\nu)}} & \frac{1}{\sqrt{1+X^2}}
& 0 \\ \frac{Y}{\sqrt{1+Y^2}} e^{-i\varphi^{(\nu)}} & 0 & 0 & \frac{1}{\sqrt{1+
Y^2}} \end{array}\right) 
\;, 
\end{equation}

\vspace{-0.2cm}

\ni where 

\vspace{-0.3cm}

%rownanie 10
\begin{eqnarray}
Y & = & \frac{M_{11}^{(\nu)} - M^{(\nu)}_{44}}{2|M_{14}^{(\nu)}|} + \sqrt{1 + 
\left(\frac{M_{11}^{(\nu)} - M^{(\nu)}_{44}}{2|M_{14}^{(\nu)}|}\right)^2}\;,
\nonumber \\ & & \nonumber \\
X & = & \frac{M_{22}^{(\nu)} - M^{(\nu)}_{33}}{2|M_{23}^{(\nu)}|} + \sqrt{1 + 
\left(\frac{M_{22}^{(\nu)} - M^{(\nu)}_{33}}{2|M_{23}^{(\nu)}|}\right)^2}\;, 
\end{eqnarray}

\vspace{-0.1cm}

\ni when $M^{(\nu)}_{14} = -|M^{(\nu)}_{14}| \exp i\varphi^{(\nu)}$ and $|M^{
(\nu)}_{14}| \neq 0 $, in analogy to $M^{(\nu)}_{23} = -|M^{(\nu)}_{23}| 
\exp i\varphi^{(\nu)}$ and $|M^{(\nu)}_{23}| \neq 0 $ for $\beta^{(\nu)} > 0 $.
If there is no sterile neutrino, then $ Y \rightarrow 0 $ as seen from Eq. (9),
what corresponds in the case of $ M_{11}^{(\nu)} = 0 $ to the limit 
$|M_{14}^{(\nu)}| \rightarrow 0 $ and $ M_{44}^{(\nu)} \rightarrow 0 $ with $
M_{44}^{(\nu)}/|M_{14}^{(\nu)}| \rightarrow \infty $.

 The neutrino states $\nu_\alpha \equiv \nu_e\,,\,\nu_\mu\,,\,\nu_\tau\,,\,
\nu_s $ (of which $\nu_e\,,\,\nu_\mu\,,\,\nu_\tau $ denote the familiar 
observed neutrino weak--interaction states, while $\nu_s $ stands for their 
unobservable sterile partner) are related to neutrino mass states $\nu_J \equiv
\nu_1\,,\,\nu_2\,,\,\nu_3\,,\,\nu_4 $ through the four--dimensional unitary 
transformation

\vspace{-0.2cm}

%rownanie 11
\begin{equation}
\nu_\alpha = \sum_J V^*_{J\,\alpha}\,\nu_J
\end{equation}

\vspace{-0.1cm}

\ni with $ \left(V^*_{J\,\alpha}\right)= \left(V_{\alpha\,J} \right)^\dagger $.
Here,

\vspace{-0.2cm}

%rownanie 12
\begin{equation}
V_{\alpha\,J} \equiv \sum_K U^{(\nu)\,*}_{K\,\alpha}U^{(e)}_{K\,J} = 
\sum_k U^{(\nu)\,*}_{k\,\alpha}U^{(e)}_{k\,J} + U^{(\nu)\,*}_{4\alpha}\delta_{
4J}
\end{equation}

\vspace{-0.1cm}

\ni with the charged--lepton diagonalizing matrix $\left(U^{(e)}_{ij}\right)$
as given in Eq. (6) and

\vspace{-0.1cm}

%rownanie 13 
\begin{equation}
U^{(e)}_{i4} = 0\;,\;\;U^{(e)}_{4j} = 0\;,\;\;U^{(e)}_{44} = 1\;.
\end{equation}

\vspace{-0.1cm}

\ni The latter equations follow from the fact that charged leptons get no 
sterile partner. Thus,

\vspace{-0.1cm}

%rownanie 14
\begin{equation}
V_{\alpha\,j} = \sum_k U^{(\nu)\,*}_{k\,\alpha}U^{(e)}_{k\,j}\;,\; 
V_{\alpha\,4} = U^{(\nu)\,*}_{4\,\alpha}\; .
\end{equation}

\vspace{-0.1cm}

\ni Of course, the $ 4\times 4 $ unitary matrix $\left(V_{\alpha\,J}\right)$ 
is the four--dimensional lepton counterpart of the familiar \CKM  matrix for 
quarks. The charged leptons $ e^-\,,\,\mu^-\,,\,\tau^-$ (with diagonalized mass
matrix) are here counterparts of the up quarks $u\,,\,c\,,t $ (with diagonal\-%
ized mass matrix).

 From Eqs. (12) as well as (9) and (6) we can calculate the matrix elements
$ V_{\alpha\,J} $ in the lowest (quadratic) perturbative order in $\alpha^{(e)}
/\mu^{(e)}$. The result reads (we write for convenience $\alpha = I = 1,2,3,4
$):

\vspace{-0.1cm}

%rownanie 15
\begin{eqnarray}
V_{11} & = & \left[1 - \frac{2}{841}\left(\frac{\alpha^{(e)}}{m_\mu}\right)^2
\right]\frac{1}{\sqrt{1+Y^2}}\;, \nonumber \\
V_{22} & = & \left[1 - \frac{2}{841}\left(\frac{\alpha^{(e)}}{m_\mu}\right)^2 
- \frac{96}{841}\left(\frac{\alpha^{(e)}}{m_\tau}\right)^2 - \frac{8\sqrt{3}
}{29}\frac{\alpha^{(e)}}{m_\tau}\,X\,e^{i(\varphi^{(\nu)}-\varphi^{(e)})}
\right]\frac{1}{\sqrt{1+X^2}}\;, \nonumber \\
V_{33} & = & \left[1 - \frac{96}{841}\left(\frac{\alpha^{(e)}}{m_\tau}\right)^2
- \frac{8\sqrt{3}}{29}\frac{\alpha^{(e)}}{m_\tau}\,X\,e^{-i(\varphi^{(\nu)}-
\varphi^{(e)})}\right]\frac{1}{\sqrt{1+X^2}}\;, \nonumber \\
V_{12} & = & \frac{2}{29} \frac{\alpha^{(e)}}{m_\mu}\,\frac{1}{\sqrt{1+Y^2}}
\,e^{i\varphi^{(e)}}\;\;,\;\;V_{21} = -\frac{2}{29} \frac{\alpha^{(e)}}{
m_\mu}\,\frac{1}{\sqrt{1+X^2}}\,e^{-i\varphi^{(e)}}\;, \nonumber \\
V_{23} & = & \left\{\left[1 - \frac{96}{841}\left(\frac{\alpha^{(e)}}{m_\tau}
\right)^2\right]X\,e^{i\varphi^{(\nu)}} + \frac{8\sqrt{3}}{29}
\frac{\alpha^{(e)}}{m_\tau}\,e^{i\varphi^{(e)}}\right\}\frac{1}{\sqrt{1+X^2}}
\;, \nonumber \\ 
V_{32} & = & \left\{\left[-1 + \frac{2}{841}\left(\frac{\alpha^{(e)}}{m_\mu}
\right)^2 + \frac{96}{841}\left(\frac{\alpha^{(e)}}{m_\tau}\right)^2\right]
X\,e^{-i\varphi^{(\nu)}} - \frac{8\sqrt{3}}{29}\frac{\alpha^{(e)}}{m_\tau}\,
e^{-i\varphi^{(e)}}\right\}\frac{1}{\sqrt{1+X^2}}\;, \nonumber \\
V_{13} & = & 0 \;\;\;,\;\;\;V_{31} = \frac{2}{29}\frac{\alpha^{(e)}}{m_\mu}
\frac{X}{\sqrt{1+X^2}}\,e^{-i(\varphi^{(\nu)}+\varphi^{(e)})}\;, \nonumber\\
V_{14} & = & \frac{Y}{\sqrt{1+Y^2}}\,e^{i\varphi^{(\nu)}}\;\;\;,\;\;\;
V_{41} = \left[-1 + \frac{2}{841}\left(\frac{\alpha^{(e)}}{m_\mu}\right)^2
\right]\frac{Y}{\sqrt{1+Y^2}}\,e^{-i\varphi^{(\nu)}}\;, \nonumber \\
V_{24} & = & 0\;\;,\;\;V_{42} = -\frac{2}{29}\frac{\alpha^{(e)}}{m_\mu}\,\frac{
Y}{\sqrt{1+Y^2}}\,e^{-i(\varphi^{(\nu)}-\varphi^{(e)})}\;, \nonumber \\
V_{34} & = & 0\;\;,\;\;V_{43} = 0\;\;,\;\;V_{44} = \frac{1}{\sqrt{1+Y^2}}
\;.
\end{eqnarray}

\vfill\eject

\vspace{0.3cm}

\ni {\bf 3. Neutrino oscillations}

\vspace{0.3cm}

 Once knowing the elements (15) of the lepton \CKM matrix, we are able to 
calculate the probabilities of neutrino oscillations $\nu_\alpha \rightarrow 
\nu_\beta $ (in the vacuum), making use of the familiar formulae:

%rownanie 16
\begin{equation}
P(\nu_\alpha \rightarrow \nu_\beta) = |\langle\nu_\beta |\nu_\alpha(t)\rangle
|^2 = \sum_{K\,L}V_{L\,\beta}V^*_{L\,\alpha}V^*_{K\,\beta}V_{K\,\alpha} 
\exp\left(i\frac{m^2_{\nu_L}-m^2_{\nu_K}}{2|\vec{p}|}\,t\right)\;,
\end{equation}

\ni where $\nu_\alpha(0) = \nu_\alpha $, $\langle\nu_\alpha| =\langle 0 |
\nu_\alpha $ and $\langle\nu_\beta |\nu_\alpha \rangle = \delta_{\beta\,\alpha}
$. Here, as usual, $t/|\vec{p}| = L/E\,\;( c = 1 = \hbar)$ and this is equal 
to $ 4\times 1.2663 L/E $ if $m^2_{\nu_L} - m^2_{\nu_K}\;,\; L $ and $ E $ are
measured in eV$^2$, m and MeV, respectively. Of course, $ L $ is the source--%
detector distance (the baseline). In the following, it will be convenient to 
denote $ x_{LK} = 1.2663(m^2_{\nu_L} - m^2_{\nu_K}) L/E $ and use the identity
$ \cos 2x_{LK} = 1 - 2\sin^2 x_{LK}$.

 From Eqs. (16) and (15) we deduce by explicit calculations the following 
neutrino oscillation formulae valid in the lowest (quadratic) perturbative 
order in $\alpha^{(e)}/\mu^{(e)}$:

%rownanie 17, 18, 19
\begin{eqnarray}
\lefteqn{P\left(\nu_e \rightarrow \nu_\mu  \right) = \frac{16}{841} 
\left(\frac{\alpha^{(e)}}{m_\mu}\right)^2 } \nonumber \\ 
& & \times \left[\frac{1}{(1+X^2)(1+Y^2)}\left(\sin^2 x_{21}+X^2 \sin^2 x_{31}
+Y^2 \sin^2 x_{24}+ X^2 Y^2 \sin^2 x_{34}\right)\right. \nonumber \\ 
& & \;\;\;-\left. \frac{X^2}{(1+X^2)^2}\sin^2 x_{32} - \frac{Y^2}{(1+Y^2)^2}
\sin^2 x_{41} \right]\;,\\
\lefteqn{P\left(\nu_e \rightarrow \nu_\tau \right) = \frac{16}{841} \left(
\frac{\alpha^{(e)}}{m_\mu}\right)^2 \frac{X^2}{(1+X^2)^2} \sin^2 x_{32}\;,} \\
\lefteqn{P\left(\nu_\mu \rightarrow \nu_\tau\right) = } \nonumber \\
& & \left\{ \left[1 -\frac{4}{841}\left(\frac{\alpha^{(e)}}{m_\mu}\right)^2 
\right]\frac{4 X^2}{(1+X^2)^2} + \frac{64\sqrt{3}}{29}\,
\frac{\alpha^{(e)}}{m_\tau}\,\frac{X(1-X^2)}{(1+X^2)^2}\cos \left(
\varphi^{(\nu)} - \varphi^{(e)}\right)\right.  \nonumber \\
& + & \left.\frac{768}{841} \left(\frac{\alpha^{(e)}}{m_\tau}\right)^2
\left[ \left(1-X^2\right) - 4X^2 \cos^2\left(\varphi^{(\nu)} - \varphi^{(e)}
\right)\right] \frac{1}{(1+X^2)^2} \right\} \sin^2 x_{32}
\end{eqnarray}

\ni for the appearance experiments (with the appearance modes of $\nu_\mu $,
$\nu_\tau $), and

%rownanie 20, 21, 22
\begin{eqnarray}
P\left(\nu_e \rightarrow \nu_e\right) & = & 1 - P\left(\nu_e \rightarrow 
\nu_\mu\right) - P\left(\nu_e \rightarrow \nu_\tau\right) \nonumber \\
& & - \left[1 - \frac{4}{841}\left(\frac{\alpha^{(e)}}{m_\mu}\right)^2 \right]
\frac{4 Y^2}{(1+Y^2)^2} \sin^2 x_{41} \;, \\
P\left(\nu_\mu \rightarrow \nu_\mu\right) & = & 1 - P\left(\nu_\mu \rightarrow
\nu_e\right) - P\left(\nu_\mu \rightarrow \nu_\tau\right)  \nonumber \\
& & - \frac{16}{841}\left(\frac{\alpha^{(e)}}{m_\mu}\right)^2 
\frac{Y^2}{(1+Y^2)^2} \sin^2 x_{41} \;,  \\ P\left(\nu_\tau \rightarrow 
\nu_\tau\right) & = & 1 - P\left(\nu_\tau\rightarrow \nu_e \right) - 
P\left(\nu_\tau \rightarrow \nu_\mu\right) 
\end{eqnarray}

\ni for the disappearance experiments (with disappearance modes of $\nu_e $,
$\nu_\mu $, $\nu_\tau $). Of all these formulae, only Eqs. (18) and (19) have
two--family forms.

 Notice that Eqs. (17)---(19) are invariant under the simultaneous substitution
$\varphi^{(\nu)} \rightarrow -\varphi^{(\nu)}$ and $\varphi^{(e)} \rightarrow
-\varphi^{(e)} $, what means their invariance under the replacement $ V_{K\,
\alpha} \rightarrow V^*_{K\,\alpha}$. This implies through Eq. (16) that $ P(
\nu_\beta \rightarrow \nu_\alpha) = P(\nu_\alpha \rightarrow \nu_\beta)$. Thus,
our neutrino oscillation formulae, valid in the lowest (quadratic) perturbative
order in $\alpha^{(e)}/\mu^{(e)}$, preserve T reversal and, by CPT invariance, 
also CP reflection, though the matrix $\left(V_{\alpha\,J}\right)$ is here not 
real (the effect of nonreal $\left(V_{\alpha\,J}\right)$ appears in higher 
perturbative orders and then spoils CP conservation). Note that CP conservat%
ion, when it works, and CPT invariance imply $ P(\nu_\alpha \rightarrow 
\nu_\beta) = P(\bar{\nu}_\alpha \rightarrow \bar{\nu}_\beta)$ and $P(\nu_\alpha
\rightarrow \nu_\beta) = P(\bar{\nu}_\beta \rightarrow \bar{\nu}_\alpha)$, 
respectively.

 Obviously, Eqs. (20)---(22) tell us that

%rownanie 23, 24, 25
\begin{eqnarray}
P\left(\nu_e \rightarrow \nu_s\right) & = & \left[1 - \frac{4}{841}\left(
\frac{\alpha^{(e)}}{m_\mu}\right)^2 \right] \frac{4 Y^2}{(1+Y^2)^2} 
\sin^2 x_{41} \;, \\
P\left(\nu_\mu \rightarrow \nu_s\right) & = & \frac{16}{841}\left(
\frac{\alpha^{(e)}}{m_\mu}\right)^2 \frac{Y^2}{(1+Y^2)^2} \sin^2 x_{41} \;, \\
P\left(\nu_\tau \rightarrow \nu_s\right) & = & 0\;,
\end{eqnarray}

\ni for the "appearance" modes of unobservable sterile neutrino $\nu_s $.

 If there is no sterile neutrino $\nu_s $, then $ Y = 0 $ wherever it appears 
in Eqs. (17)---(24). On the other hand, if $(\alpha^{(e)}/m_\mu)^2 \rightarrow
0 $, then

%rownanie 26
\begin{equation}
P\left(\nu_\mu \rightarrow \nu_\tau\right) \rightarrow \frac{4 X^2}{(1+X^2)^2}
\sin^2 x_{32}
\end{equation}

\ni and 

%rownanie 27
\begin{equation}
P\left(\nu_e \rightarrow \nu_s\right) \rightarrow \frac{4 Y^2}{(1+Y^2)^2} 
\sin^2 x_{41}\;,
\end{equation}

\ni while $ P\left(\nu_e \rightarrow \nu_\mu \right) \rightarrow 0 $, $P\left(
\nu_e \rightarrow \nu_\tau\right) \rightarrow 0 $ and $P\left(\nu_\mu 
\rightarrow \nu_s \right) \rightarrow 0 $. Thus,

%rownanie 28
\begin{equation}
P\left(\nu_e \rightarrow \nu_e\right) \rightarrow 1 - \frac{4 Y^2}{(1+Y^2)^2} 
\sin^2 x_{41}
\end{equation}

\ni and

%rownanie 29
\begin{equation}
P\left(\nu_\mu \rightarrow \nu_\mu\right) \rightarrow 1 - \frac{4 X^2}{(1
+X^2)^2} \sin^2 x_{32}\;,
\end{equation}

\ni but $ P\left(\nu_\tau \rightarrow \nu_\tau\right) \rightarrow 1 $.

 Concluding, we can see that the practically decoupled oscillations $\nu_\mu 
\rightarrow \nu_\tau $ and $\nu_e \rightarrow \nu_s $ play in the framework of 
our neutrino "texture" an exceptional, dominating role. If there is no sterile 
neutrino $\nu_s $ {\it i.e.}, $ Y = 0 $, then such a role is played only by $
\nu_\mu \rightarrow \nu_\tau $.

 Note that in Eqs. (17)---(24) $\sin^2 x_{LK} \rightarrow 1/2 $ if $ x_{LK}/\pi
\rightarrow \infty $, since the source and detector have always finite 
extensions over which $\sin^2 x_{LK}$ ought to be averaged (with respect to 
their distance $ L $). If $ 2 x_{LK}/\pi \rightarrow 0 $, then $\sin^2 x_{LK}
=\rightarrow x^2_{LK} \rightarrow 0 $.

\vspace{0.3cm}

\ni {\bf 4. Information from atmospheric neutrinos}

\vspace{0.3cm}

 The atmospheric neutrino experiments seem to indicate that there is a deficit 
of atmospheric $\nu_\mu$'s, caused by the neutrino oscillations corresponding 
to disappearance modes of $\nu_\mu $. These result in the survival probability
which, if analized in two--family form

%rownanie 30
\begin{equation}
P\left(\nu_\mu \rightarrow \nu_\mu\right) = 1 - \sin^2 2\theta_{\rm atm}
\sin^2 \left(1.27 \Delta m^2_{\rm atm}\, L/E \right)\;,
\end{equation}

\ni leads to

%rownanie 31
\begin{equation}
\sin^2 2\theta_{\rm atm} = O(1) \sim 0.8\;\;{\rm to}\;\; 1
\end{equation}

\ni and

%rownanie 32
\begin{equation}
\Delta m^2_{\rm atm} \sim (0.03\;\;{\rm to}\;\; 1)\times 10^{-2}\;\;{\rm eV}^2
\end{equation}

\ni with the preferable value $\Delta m^2_{\rm atm} \sim 0.5\times 10^{-2}\;\;
{\rm eV}^2$ [4,5]. It is usually suggested that, practically, the oscillations
$\nu_\mu \rightarrow \nu_\tau $ alone are responsible for such a deficit. Then,

%\vspace{0.1cm}

%rownanie 33
\begin{equation}
P\left(\nu_\mu \rightarrow \nu_\tau\right) = \sin^2 2\theta_{\rm atm}
\sin^2 \left(1.27 \Delta m^2_{\rm atm}\, L/E \right)\;.
\end{equation}

%\vspace{0.1cm}

 The last suggestion, neglecting the disapearance mode $\nu_\mu \rightarrow 
\nu_e $, is consistent with the negative result of CHOOZ long--baseline reactor
experiment that finds no evidence for neutrino oscillations corresponding to 
the disappearance modes of $\bar{\nu}_e $, in particular $\bar{\nu_e} 
\rightarrow \bar{\nu}_\mu $ [6]. The region of $\sin^2 2\theta_{\rm atm}$ 
and $\Delta m^2_{\rm atm}$ indicated by Super--Kamiokande atmospheric--neutrino
experiment [Eqs. (31) and (32)] lies, in fact, inside the region of $\sin^2 2
\theta_{\rm CH}$ and $\Delta m^2_{\rm CH}$ excluded by CHOOZ (where $ P\left(
\bar{\nu_e} \rightarrow \bar{\nu}_e \right) = 1 $). 

 This important message from CHOOZ experiment, restricting the strength of mix%
ing $\nu_\mu $ with $\nu_e $, leaves {\it a priori} three options for mixing $
\nu_\mu $ with $\nu_\tau $ and $\nu_s $: $\nu_\mu $ mixes {\it dominantly} with
$\nu_\tau $ (while $\nu_e $ with $\nu_s $), or with $\nu_s $ (while $\nu_e $ 
with $\nu_\tau$), or with both $\nu_\tau $ and $\nu_s $ (while $\nu_e $ does 
not mix). Of these three options, our neutrino "texture" chooses the first due 
to the assumptions $ M^{(\nu)}_{12} = 0$ and $ M^{(\nu)}_{24} = 0 $, supplement%
ed by $ M^{(\nu)}_{13} = 0 $ and $ M^{(\nu)}_{34} = 0 $. Note that the second 
option would correspond to different assumptions, $ M^{(\nu)}_{12} = 0 $ and $ 
M^{(\nu)}_{23} = 0 $, supplemented by $ M^{(\nu)}_{14} = 0 $ and $ M^{(\nu)}_{
34} = 0 $; eventually, in the case of the third option there would be $ M^{(\nu
)}_{12} = 0 $, supplemented by $ M^{(\nu)}_{13} = 0 $, $ M^{(\nu)}_{14} = 0 $ 
and $ M^{(\nu)}_{34} = 0 $.

 We can see from the experimental atmospheric--neutrino estimates (30) and (33)
that they correspond exactly to our neutrino oscillation formulae (29) and
(26), respectively. Hence, we can infer that

%\vspace{0.1cm}
		
%rownanie 34
\begin{equation}
\frac{4 X^2}{(1+X^2)^2} = \sin^2 2\theta_{\rm atm}\;\;,\;\;
|m^2_{\nu_3} - m^2_{\nu_2}| = \Delta m^2_{\rm atm}
\end{equation}

%\vspace{0.1cm}

\ni and so, with the Super--Kamiokande figures (31) and (32), we can take 

%\vspace{0.1cm}

%rownanie35
\begin{equation} 
\frac{4 X^2}{(1+X^2)^2} \sim 0.8\;\;{\rm to}\;\; 1
\end{equation}

\ni and

%rownanie 36
\begin{equation}
|m^2_{\nu_3} - m^2_{\nu_2}| \sim 5 \times 10^{-3}\;\;{\rm eV}^2
\end{equation}

%\vspace{0.1cm}

\ni as two neutrino inputs.

 From the input (35) we evaluate the limits

%\vspace{0.1cm}

%rownanie 37
\begin{equation}
X \sim 0.618\;\;{\rm to}\;\; 1.
\end{equation}

%\vspace{0.1cm}

\ni However, it is not difficult to see that (because the difference (36) is 
kept fixed) the limit of $ X\, \rightarrow \,1 $ is singular in the sense that 
then $\mu^{(\nu)}\, \rightarrow\, 0 $ and $\beta^{(\nu)}\, \rightarrow\,
\infty $ with $\,\mu^{(\nu)}\beta^{(\nu)}\, \rightarrow (5/20.9)\times 10^{-3}
\;{\rm eV}^2\, $ (and so, $ M^{(\nu)}_{22}\, \rightarrow\, 0 $, $ M^{(\nu)}_{
33}\, \rightarrow\, 0 $ and $ |M^{(\nu)}_{23}|\, \rightarrow\, \infty $ with 
$ 2(M^{(\nu)}_{22} + M^{(\nu)}_{33})|M^{(\nu)}_{23}| \rightarrow 5\times 
10^{-3}\;\;{\rm eV}^2 $). Then, $ m_{\nu_2,\,\nu_3} =\rightarrow \mp 0.478\;
\beta^{(\nu)} \rightarrow \mp\infty $. We will restrict, therefore, the range 
in the input (35) to 

%\vspace{0.1cm}

%rownanie 38
\begin{equation}
\frac{4 X^2}{(1+X^2)^2} \sim 0.8\;\;{\rm to}\;\; 1 - 10^{-6}\;,
\end{equation}

%\vspace{0.1cm}

\ni where the particular upper limit is chosen as an illustration. Hence,

%\vspace{0.1cm}

%rownanie 39
\begin{equation}
X \sim 0.618\;\;{\rm to}\;\; 0.999\;.
\end{equation}

%\vspace{0.1cm}

\ni In such a case, the second Eq. (10) leads to

%\vspace{0.1cm}

%rownanie 40 
\begin{equation}
\frac{M^{(\nu)}_{22}-M^{(\nu)}_{33}}{2|M^{(\nu)}_{23}|} = \frac{X^2 - 1}{2X}
\sim - (0.500\;\;{\rm to}\;\;0.001)
\end{equation}

%\vspace{0.1cm}

\ni (respectively). On the other hand, the mass matrix $\left( M^{(\nu)}_{ij}
\right)$ as given in Eq. (1) implies that

%\vspace{0.1cm}

%rownanie41
\begin{equation}
\frac{M^{(\nu)}_{22}-M^{(\nu)}_{33}}{2|M^{(\nu)}_{23}|} = -20.3 
\frac{\mu^{(\nu)}}{\beta^{(\nu)}}\,.
\end{equation}

%\vspace{0.1cm}

\ni Thus, from Eqs. (40) and (41) we get

%\vspace{0.1cm}

%rownanie 42
\begin{equation}
\frac{\mu^{(\nu)}}{\beta^{(\nu)}} \sim 0.0246\;\;{\rm to}\;\;0.0000492\;,\;
\frac{\beta^{(\nu)}}{\mu^{(\nu)}} \sim 40.7\;\;{\rm to}\;\;20300
\end{equation}

\ni (respectively).

Then, making use of Eq. (1) for the mass matrix $\left( M^{(\nu)}_{ij}\right)$,
we obtain from the second Eq. (8)

\vspace{-0.1cm}

%rownanie 43
\begin{eqnarray}
m_{\nu_2,\,\nu_3} & = & \left[10.9 \mp 0.478 \frac{\beta^{(\nu)}}{\mu^{(\nu)}}
\sqrt{1 + \left(20.3\frac{\mu^{(\nu)}}{\beta^{(\nu)}} \right)^2} \right]
\mu^{(\nu)} \nonumber \\
& = & \left\{\begin{array}{r} -(10.8\;\;{\rm to}\;\;9700)\mu^{(\nu)} \\ 
(32.7\;\;{\rm to}\;\;9730)\mu^{(\nu)} \end{array} \right.
\end{eqnarray}

\ni (respectively). Therefore,

\vspace{-0.1cm}

%rownanie 44
\begin{equation}
m^2_{\nu_3} - m^2_{\nu_2} = (951\;\;{\rm to}\;\;425000)\mu^{(\nu)\,2}\;. 
\end{equation}

\ni Hence, using the input (36), we evaluate

\vspace{-0.1cm}

%rownanie 45
\begin{equation}
\mu^{(\nu)} \sim (0.00229\;\;{\rm to}\;\;0.000108)\;{\rm eV}
\end{equation}

\ni (respectively). With the values (45) of $\mu^{(\nu)} $, Eqs. (43) lead to

\vspace{-0.1cm}

%rownanie 46 
\begin{equation}
m_{\nu_2} \sim -(0.0247\;\;{\rm to}\;\;1.05)\;{\rm eV}\;\;,\;\;m_{\nu_3} \sim 
(0.0749\;\;{\rm to}\;\;1.05)\;{\rm eV} 
\end{equation}

\ni (respectively). Here, the minus sign at $ m_{\nu_2}$ is phenomenologically 
irrelevant in relativistic dynamics ({\it cf.} Dirac equation). Note that

\vspace{-0.1cm}

%rownanie 47
\begin{equation}
m_{\nu_2}^2 \sim (0.000611\;\;{\rm to}\;\;1.11)\;{\rm eV}^2\;\;,\;\;
m_{\nu_3}^2 \sim (0.0561\;\;{\rm to}\;\;1.11)\;{\rm eV}^2 \;,
\end{equation}

\ni where $ m^2_{\nu_3} - m^2_{\nu_2} \sim 0.005\;{\rm eV}^2 $. Finally, from 
Eqs. (42) and (45) we evaluate

\vspace{-0.1cm}

%rownanie 48
\begin{equation}
\beta^{(\nu)} \sim (0.0933\;\;{\rm to}\;\;2.20)\;{\rm eV} \;.
\end{equation}

 To summarize the above discussion of atmospheric neutrinos, we can say that 
the Super--Kamiokande experiment seems to transmit to us an important message 
about strong mixing of $\nu_\mu $ and $\nu_\tau $ and their rather weak mixing 
with $\nu_e $. Such a situation is predicted just in the case of our neutrino 
"texture" with $ X = O(1) $ and small $\left( \alpha^{(e)}/\mu^{(e)}\right)^2 
\rightarrow 0 $, the latter value being motivated by our excellent zero--order 
prediction of $ m_\tau $. This conclusion is independent of whether the 
sterile neutrino $\nu_s $ exists or does not exist in our neutrino "texture". 
The predicted neutrino masses are $|m_{\nu_2}| \sim (0.02\;\;{\rm to}\;\;1)$ 
eV and $ m_{\nu_3} \sim (0.07\;\;{\rm to}\;\;1)$ eV with $ m_{\nu_3}^2 - 
m_{\nu_2}^2 \sim 5 \times 10^{-3}\;{\rm eV}^2 $, where the upper limit $ X = 
0.999 $ is considered for $ X < 1 $. For $ X $ still nearer to 1, $ m_{\nu_2}$ 
and $ m_{\nu_3}$ increase further.

\vspace{0.3cm}

\ni {\bf 5. Information from solar neutrinos}

\vspace{0.3cm}

 As is well known, the solar neutrino experiments demonstrate a deficit of 
solar $\nu_e$'s reaching the Earth, that seems to be caused by neutrino oscil%
lations corresponding to disappearance modes of $\nu_e $. These modes result in
the survival probability, usually analized in two--family form

\vspace{-0.15cm}

%rownanie 49 
\begin{equation}
P\left(\nu_e \rightarrow \nu_e\right) = 1 - \sin^2 2\theta_{\rm sol}
\sin^2 \left(1.27 \Delta m^2_{\rm sol}\, L/E \right)\;.
\end{equation}

\ni Here, the oscillation amplitude $\sin^2 2\theta_{\rm sol}$ is likely to be 
enhanced to a new $(\sin^2 2\theta_{\rm sol})_{\rm matter}$ by the resonant MSW
mechanism [7] in the Sun matter (dependent on values of $\sin^2 2\theta_{\rm 
sol}$ and $\Delta m^2_{\rm sol}$), though the vacuum mechanism is still not 
excluded. According to recent estimations [8], in the first case there are two 
solutions with

\vspace{-0.15cm}

%rownanie 50 
\begin{equation}
\sin^2 2\theta_{\rm sol} \sim 8 \times 10^{-3}\;\;,\;\;\Delta m^2_{\rm sol}
\sim 5 \times 10^{-6}\;{\rm eV}^2
\end{equation}

\ni and

\vspace{-0.15cm}

%rownanie 51 
\begin{equation}
\sin^2 2\theta_{\rm sol} \sim 0.6\;\;,\;\;\Delta m^2_{\rm sol} \sim 1.6 
\times 10^{-5}\;{\rm eV}^2\;,
\end{equation}

\ni while in the second case

\vspace{-0.15cm}

%rownanie 52 
\begin{equation}
\sin^2 2\theta_{\rm sol} \sim 0.65\, - \,1\;\;,\;\;\Delta m^2_{\rm sol} \sim 
(5\, - \,8) \times 10^{-11}\;{\rm eV}^2 \;.
\end{equation}

\ni If the disappearance mode $\nu_e \rightarrow \nu_s $ dominates, then from 
the two MSW solutions only the first survives. From the above three solutions, 
the first is considered as most favorable.

 Note that the two values $\Delta m^2_{\rm sol} = O(10^{-5}\;{\rm eV}^2)$ 
indicated in the first case as well as the value $\Delta m^2_{\rm sol} = O(10^{
-10}\;{\rm eV}^2)$ allowed in the second are situated --- consistently --- much
below the region excluded for disappearance modes of $\bar{\nu}_e $ by CHOOZ 
experiment, where $\Delta m^2_{\rm CH} \stackrel{>}{\sim} 0.9 \times 10^{-3}\;{
\rm eV}^2 $ (at the 90\% confidence level) [6].

 We can see from the experimental solar--neutrino estimate (49) that it may be
related to our neutrino oscillation formula (28) (both being likely enhanced by
the MSW mechanism). Then, only the disappearance mode $\nu_e \rightarrow \nu_s 
$, described by the formula (27), contributes to the rhs of Eq. (20). In such 
a case we get with the use of figures (50) two other neutrino inputs

\vspace{-0.15cm}

%rownanie 53 
\begin{equation}
\frac{4 Y^2}{(1+Y^2)^2} = \sin^2 2\theta_{\rm sol} \sim 8\times 10{-3}
\end{equation}

\vspace{-0.1cm}

\ni and

\vspace{-0.15cm}

%rownanie 54 
\begin{equation}
|m^2_{\nu_4} - m^2_{\nu_1}| = \Delta m^2_{\rm sol} \sim 5 \times 10^{-6}\;
{\rm eV}^2 \;,
\end{equation}

\ni if the first MSW solution is taken into account. The input (53) gives $ Y 
\sim 0.045 $ (and $ Y^2 \sim 2.0\times 10^{-3}$). 

 From the first Eq. (10) with $ M^{(\nu)}_{11} = 0 $ we obtain 

\vspace{-0.15cm}

%rownanie 55
\begin{equation}
Y \simeq \frac{|M^{(\nu)}_{14}|}{M^{(\nu)}_{44}}
\end{equation}

\ni under the assumption that $ M^{(\nu)\,2}_{44} \gg \left(2|M^{(\nu)}_{14}|
\right)^2 $. Thus, $\left(|M^{(\nu)}_{14}|/M^{(\nu)}_{44}\right)^2 \sim 0.0020 
$ with the input (53).

 On the other hand, from the first Eq. (8) with $ M^{(\nu)}_{11} = 0 $

\vspace{-0.15cm}

%rownanie 56
\begin{equation}
m_{\nu_1} \simeq -\frac{|M^{(\nu)}_{14}|^2}{M^{(\nu)}_{44}}\;\;,\;\;
m_{\nu_4} \simeq M^{(\nu)}_{44} + \frac{|M^{(\nu)}_{14}|^2}{M^{(\nu)}_{44}}\;,
\end{equation}

\ni when $M^{(\nu)\,2}_{44} \gg \left(2 |M^{(\nu)}_{14}|\right)^2 $. Thus, $ 
M^{(\nu)\,2}_{44} \sim 5\times 10^{-6}\;{\rm eV}^2$ with the input (54), and 
so, $ |M^{(\nu)}_{14}|^2 \sim 1.0\times 10^{-8}\;{\rm eV}^2$ due to the input 
(53). Hence,

\vspace{-0.15cm}

%rownanie 57
\begin{equation}
m_{\nu_1} \sim -4.5\times 10^{-6}\;{\rm eV}\;\;,\;\;m_{\nu_4} \sim 2.2\times 
10^{-3}\;{\rm eV}\;.
\end{equation}

\ni Here again, the minus sign at $m_{\nu_1}$ is phenomenologically irrelevant.

 We can see from Eq. (20) that its approximate form (28) valid for $\left(
\alpha^{(e)}/m_\mu\right)^2 \rightarrow 0 $, applied here, works very well 
even for the central value of $\left( \alpha^{(e)}/m_\mu\right)^2 $, because 
then

\vspace{-0.15cm}

%rownanie 58
\begin{eqnarray}
\frac{4 Y^2}{(1+Y^2)^2} \sim 8\times 10^{-3}\;\;{\rm large}\;{\it versus}\;\;
\frac{16}{841}\left(\frac{\alpha^{(e)}}{m_\mu}\right)^2 \sim 2.5\times 10^{-4}
\;.
\end{eqnarray}

\ni In fact, due to Eq. (5),

\vspace{-0.15cm}

%rownanie 59
\begin{equation}
0 \leq \frac{16}{841}\left(\frac{\alpha^{(e)}}{m_\mu}\right)^2 \leq 6.2\times 
10^{-4}\;,
\end{equation}

\ni where the central value is $2.5 \times 10^{-4} $.

 The small value (59) is the main reason, why in the case of no sterile 
neutrino $\nu_s $ our neutrino oscillation formula (20) cannot be compared with
the experimental estimate (49). Indeed, in this case it assumes the form

\vspace{-0.15cm}

%rownanie 60
\begin{equation}
P\left(\nu_e \rightarrow \nu_e\right) = 1 - \frac{16}{841}\left(\frac{
\alpha^{(e)}}{m_\mu}\right)^2 \frac{1}{1+X^2}\left(\sin^2 x_{21} + X^2\sin^2 
x_{31} \right)\;.
\end{equation}

 To conclude the above discussion of solar neutrinos, we can claim that, in the
framework of our "texture", only the disappearance mode of $\nu_e $ to the 
sterile neutrino $\nu_s $ (likely to be enhanced by the MSW mechanism in the 
Sun matter) can be responsible for the observed deficit of solar $\nu_e$'s. 
The predicted neutrino masses are $|m_{\nu_1}| \sim 2\times 10^{-6}$ eV and $ 
m_{\nu_4} \sim 2\times 10^{-3}$ eV with $m^2_{\nu_4} - m^2_{\nu_1} \sim 5
\times 10^{-6}\;{\rm eV}^2$.

 Our last remark concerns the LSND experiment that seems to detect $\nu_\mu 
\rightarrow \nu_e $ oscillations by observing the appearance of $\nu_e $ 
originating from $\nu_\mu $ produced in $\pi^+ $ decay [9]. The observed 
excess of $\nu_e$'s, analized in terms of two--family oscillation formula, 
leads to $\sin^2 2\theta_{\rm LSND} > 4\times 10^{-4}$, in particular to 
$\sin^2 2\theta_{\rm LSND} \sim 1.5\times 10^{-3} - 1.5\times 10^{-1}$ for 
$\Delta m^2_{\rm LSND} \sim 1\;{\rm eV}^2$ ({\it cf.} [9]). This shows that, 
due to the small value (59), the LSND magnitude of $\nu_\mu \rightarrow \nu_e 
$ oscillation amplitude can hardly be explained by means of our formula (17) 
for $P(\nu_\mu \rightarrow \nu_e) = P(\nu_e \rightarrow \nu_\mu) $ . However, 
there is possibly a narrow overlap near the upper limit of the range (59), 
corresponding to the values $\sin^2 2\theta_{\rm LSND} \sim 6 \times 10^{-4}$ 
and $\Delta m^2_{\rm LSND} \stackrel{>}{\sim} 1.2\;{\rm eV}^2 $, located at 
the border of LSND allowed region (at the 95\% confidence level) [9]. 
Unfortunately, another (earlier) LSND experiment on $\bar{\nu}_\mu \rightarrow 
\bar{\nu}_e$ oscillations estimates that an analogical allowed region is much 
narrower ({\it cf.} [9]), what excludes the above potential overlap.

 Nevertheless, it may be interesting to remark that, in the case of values $
m^2_{\nu_2} \sim m^2_{\nu_3}\stackrel{>}{\sim} 1.2\;{\rm eV}^2$ corresponding
to the large $\Delta m^2_{\rm LSND} \stackrel{>}{\sim} 1.2\;{\rm eV}^2$, we can 
put $x_{21} \sim x_{31} \sim x_{24} \sim x_{34} \gg x_{32} \geq x_{41}$ 
in Eq. (17). This leads to the two--family oscillation formula

\vspace{-0.1cm}

%rownanie 61
\begin{equation}
P\left(\nu_e \rightarrow \nu_\mu\right) \simeq \frac{16}{841}\left(
\frac{\alpha^{(e)}}{m_\mu}\right)^2 \sin^2 x_{21}\;,
\end{equation}

\vspace{-0.1cm}

\ni when short--baseline experiments with $ x_{21} \stackrel{<}{\sim} \pi/2 $ 
are considered (here, $ 1+Y^2 \sim 1+0.0020 \simeq 1 $). 

%\vfill\eject

\vspace{0.3cm}

\ni {\bf 6. Outlook: an unconventional picture}

\vspace{0.3cm}

 If beside the sterile neutrino $\nu_s $ there exists also a second sterile 
neutrino mentioned in the footnote $^{\ddagger}$ , two opposite options seem to
be attractive. The first was already outlined in this footnote: such a second 
sterile neutrino might mix mainly with $\nu_\mu $, leading to a new disappear%
ance mode of $\nu_\mu $ that, if not too strong, would only slightly correct 
the effect of the dominating mode $\nu_\mu \rightarrow \nu_\tau $. In the 
second option, the roles of these two modes of $\nu_\mu $ would be inter\-%
changed: the new disappearance mode of $\nu_\mu $ would dominate the mode $
\nu_\mu \rightarrow \nu_\tau $, and so would be responsible for producing a 
near--to--maximal oscillation amplitude for $\nu_\mu \rightarrow \nu_\mu $ as 
is observed in Super--Kamiokande. In such a case, both parameters $\alpha^{(
\nu)}$ and $\beta^{(\nu)}$ might be small, perhaps zero.

 Thus, this second option would create a uniform but unconventional picture of 
neutrino oscillations, where they would be caused essentially by mixing two 
sterile neutrinos with $\nu_e$ and $\nu_\mu $, respectively. At any rate, this 
picture would be true for the solar and atmospheric neutrinos.

 Sterile neutrinos of both kinds might constitute an important part of relati%
vistic dark matter, passive with respect to all Standard--Model interactions 
(including weak interactions). In such a case, they might even be the main
constituents of matter in the Universe. Of course, sterile neutrinos would 
interact with each other and with Standard--Model active particles through 
gravitation.

%\vfill\eject

\vspace{0.6cm}

{\centerline{\bf Appendix}}

\vspace{0.4cm}

\appendix\setcounter{equation}{0}

 For the neutrino mass matrix $\left(M^{(\nu)}_{ij}\right)$ as given in Eq. (1)
we assumed that $\!\alpha^{(\nu)} = 0\! $ [Eq. (2)], what implied $\! M^{(\nu)
}_{12}\! = 0 = \!M^{(\nu)}_{21}$. As we saw, this assumption, leading to weak 
mixing of $\!\nu_e $ with $\!\nu_\mu $ and $\!\nu_\tau $, is neatly consistent 
with negative results of CHOOZ experiment~[6]. Now, we will relax such an 
extremal assumption, allowing for a nonzero but small $\alpha^{(\nu)}$, much 
smaller than the large $\beta^{(\nu)}\! = \! O(10^{-4}\!$ to 1) eV [Eq. (48)] 
responsible for near--to--maximal mixing of $\nu_\mu $ with $\nu_\tau $, just 
as suggested by results of Super--Kamiokande experiment [4,5]. Although, in our
neutrino "texture" such a relaxation is not needed to understand solar neutrino
experiments which are here reasonably explained by the mixing of $\nu_e $ with 
$\nu_s $, it may be applied to appearance experiments in the mode $\nu_\mu 
\rightarrow \nu_e $~or~$\nu_e \rightarrow \nu_\mu $.

 In this case, $\left(M^{(\nu)}_{ij}\right)$ is perturbed by the matrix

\vspace{-0.1cm}

%(A.1)
$$
\left(\delta M^{(\nu)}_{ij}\right)  = \frac{\alpha^{(\nu)}}{29}\left(
\begin{array}{ccc} 0 & 2 e^{i\,\varphi^{(\nu)}} & 0 \\ 2e^{-i\,\varphi^{(\nu)}}
& 0 & 8\sqrt{3} e^{i\,\varphi^{(\nu)}} \\ 0 & 8\sqrt{3} e^{-i\,\varphi^{(\nu)}}
& 0 \end{array} \right)\;\;.
\eqno({\rm A}.1)
$$

\ni Thus,

\vspace{-0.1cm}

%(A.2)
$$
\delta M^{(\nu)}_{12} = \frac{\alpha^{(\nu)}}{29}e^{i\,\varphi^{(\nu)}} =
\delta M^{(\nu)\,*}_{21}\;\;,\;\;\delta M^{(\nu)}_{23} = \frac{8\sqrt{3}
\alpha^{(\nu)}}{29}e^{i\,\varphi^{(\nu)}} = \delta M^{(\nu)\,*}_{32}\;,
\eqno({\rm A}.2)
$$

\vspace{-0.1cm}

\ni while

\vspace{-0.2cm}

%(A.3)
$$
M^{(\nu)}_{12} = 0 = M^{(\nu)}_{21}\;\;,\;\;M^{(\nu)}_{23} = -\frac{8\sqrt{3}
\beta^{(\nu)}}{29} e^{i\,\varphi^{(\nu)}} = M^{(\nu)\,*}_{32}\;.
\eqno({\rm A}.3)
$$

 Then, the total secular equation det$\left[M^{(\nu)} + \delta M^{(\nu)} - {\bf
1}(m_{\nu_i} + \delta m_{\nu_i})\right] = 0 $ gives in the lowest (quadratic 
or linear) perturbative order in $\alpha^{(\nu)}/\mu^{(\nu)}$ the following 
neutrino mass corrections:

\vspace{-0.2cm}

%(A.4)
\begin{eqnarray*}
\delta m_{\nu_1}\;\; & = & -\frac{|\delta M^{(\nu)}_{12}|^2 M^{(\nu)}_{33}}{m_{
\nu_2}\,m_{\nu_3}} = -\frac{|\delta M^{(\nu)}_{12}|^2}{m_{\nu_2}}\left(1 - 
\frac{ |M^{(\nu)}_{23}|}{m_{\nu_3}}\,X \right)\;, \\
\delta m_{\nu_2,\,\nu_3} & = & \pm \frac{2|\delta M^{(\nu)}_{23}||M^{(\nu)}_{23}|
}{m_{\nu_3} - m_{\nu_2}} = \pm 2|\delta M^{(\nu)}_{23}|\frac{X}{1 + X^2}\;,
\end{eqnarray*}

\vspace{-1.52cm}

\begin{flushright}
(A.4)
\end{flushright}

\vspace{0.15cm}

\ni where $|\delta M^{(\nu)}_{12}| = 0.0690 \alpha^{(\nu)}$, $|\delta M^{(\nu)
}_{23}| = 0.478 \alpha^{(\nu)}$, $\! M^{(\nu)}_{33} = 20.7 \mu^{(\nu)}$ and 
$| M^{(\nu)}_{23}| = 0.478\, \beta^{(\nu)}$.

 When the mass matrix $\!\left(M^{(\nu)}_{ij}\right)\!$ is perturbed by the 
matrix $\!\left(\delta M^{(\nu)}_{ij}\right)\!$ given in Eq. (A.1), then the 
unitary diagonalizing matrix $\!\left(U^{(\nu)}_{ij}\right)\!$ [Eq.(9)] under%
goes the perturbation $\!\left(\delta U^{(\nu)}_{ij}\right)\!$ which, after 
some calculations in the lowest perturbative order, can be written~as~ follows:

\vspace{-0.2cm}

%(A.5)
$$
\left(\delta U^{(\nu)}_{ij}\right) =
\left(\begin{array}{ccc}  0 & \frac{|\delta M^{(\nu)}_{12}|}{m_{\nu_2}\sqrt{1+
X^2}}e^{i\varphi^{(\nu)}} & -\frac{|\delta M^{(\nu)}_{12}|X}{m_{\nu_3}\sqrt{1
+X^2}}e^{2i\varphi^{(\nu)}} \\ -\frac{|\delta M^{(\nu)}_{12}|M^{(\nu)}_{33}}
{m_{\nu_2}\,m_{\nu_3}}e^{-i\varphi^{(\nu)}} & 0 & \frac{|\delta M^{(\nu)}_{23}|
X(1-X^2)}{|M^{(\nu)}_{23}|(1+X^2)^{3/2}}e^{i\varphi^{(\nu)}} \\ -\frac{|\delta
M^{(\nu)}_{12}||M^{(\nu)}_{23}|}{m_{\nu_2}\,m_{\nu_3}}e^{-2i\varphi^{(\nu)}} & 
-\frac{|\delta M^{(\nu)}_{23}|X(1-X^2)}{|M^{(\nu)}_{23}|(1+X^2)^{3/2}}
e^{-i\varphi^{(\nu)}} & 0 \end{array}\right) \eqno({\rm A}.5)
$$

\vspace{0.1cm}

\ni with $M^{(\nu)}_{33}/m_{\nu_3} = 1 - |M^{(\nu)}_{23}|X/m_{\nu_3}$ and $
m_{\nu_2} = -|m_{\nu_2}|$. Here, $Y \sim 0.0045 $ is put zero for simplicity, 
and $i\,,\,j = 1,2,3 $ (if $Y \neq 0 $, then $\delta U^{(\nu)}_{21}$ and
$\delta U^{(\nu)}_{31}$ get extra factors $1/\sqrt{1 + Y^2}$). In particular, 
the elements $\delta U^{(\nu)}_{21}$ and $\delta U^{(\nu)}_{31}$ are produced 
(in the lowest perturbative order) through the diagonalizing procedure of the 
total mass matrix $\left(M^{(\nu)}_{ij} + \delta M^{(\nu)}_{ij}\right)$ in 
the following way:

\vspace{-0.3cm}

%(A.6)
\begin{eqnarray*}
U^{(\nu)}_{21} + \delta U^{(\nu)}_{21} & = & -\frac{M^{(\nu)}_{11}
- m_{\nu_1} - \delta m_{\nu_1} }{M^{(\nu)}_{12} + \delta M^{(\nu)}_{12}} = 
\frac{\delta m_{\nu_1} }{\delta M^{(\nu)}_{12}} = -\frac{ |\delta M^{(\nu)}_{12
}|M^{(\nu)}_{33} }{ m_{\nu_2}\,m_{\nu_3}} e^{-i\varphi^{(\nu)}}\;, \\ 
U^{(\nu)}_{31} + \delta U^{(\nu)}_{31} & = & -\frac{(M^{(\nu)}_{11} - 
m_{\nu_1} - \delta m_{\nu_1})( M^{(\nu)}_{32} + \delta M^{(\nu)}_{32})}{
(M^{(\nu)}_{12} + \delta M^{(\nu)}_{12})(M^{(\nu)}_{33}- m_{\nu_1} - \delta 
m_{\nu_1})} = \frac{\delta m_{\nu_1}\,M^{(\nu)}_{32}}{\delta M^{(\nu)}_{12}
\,M^{(\nu)}_{33}} \\ & = & -\frac{|\delta M^{(\nu)}_{12}||M^{(\nu)}_{23}|}{m_{
\nu_2}\,m_{\nu_3}} e^{-2 i\varphi^{(\nu)}} 
\;,
\end{eqnarray*}

\vspace{-1.52cm}

\begin{flushright}
(A.6)
\end{flushright}

\vspace{0.3cm}

\ni where Eq. (A.4) for $\delta m_{\nu_1}$ is used, while $ U^{(\nu)}_{21} = 0$
and $ U^{(\nu)}_{31} =0 $. Of course, $M^{(\nu)}_{11} = 0 $ and so, $ m_{\nu_1}
= 0 $ for $ Y = 0 $.

 Once $\left(\delta U^{(\nu)}_{ij}\right)$ is known, the perturbation $\left(
\delta V_{ij}\right)$ of the leptonic \CKM matrix $\left(V_{ij}\right)$ [Eqs. 
(15)] can be evaluated from the definition:

\vspace{0.1cm}

%(A.7)
$$
\left(V_{ij} + \delta V_{ij}\right) = \left(U^{(\nu)}_{ik} + \delta U^{(\nu)}_{
ik}\right)^\dagger\left(U^{(e)}_{kj}\right) \;. \eqno({\rm A}.7)
$$

\vspace{0.2cm}

\ni Here, $\nu_\alpha = \sum_j \left(V_{j\alpha}^* + \delta V_{j\alpha}^*
\right) \nu_j $ with $\nu_\alpha = \nu_e\,,\,\nu_\mu\,,\,\nu_\tau $ and $\nu_j
= \nu_1\,,\,\nu_2\,,\,\nu_3 $. Thus, from Eq. (A.7) we obtain

\vspace{0.1cm}

%(A.8)
$$ 
\left(\delta V_{ij}\right) = \left(\delta U^{(\nu)}_{ik}\right)^\dagger
\left(U^{(e)}_{kj}\right) \simeq \left(\delta U^{(\nu)}_{ij}\right)^\dagger \;,
\eqno({\rm A}.8)
$$

\vspace{0.2cm}

\ni where, in the second step, terms proportional to $(\alpha^{(\nu)}/m_{
\nu_2})(\alpha^{(e)}/m_\mu)$ are neglected {\it versus} terms proportional to 
$(\alpha^{(\nu)}/m_{\nu_2})$. Thus, in this case

\vspace{0.1cm}

%(A.9)
$$
\delta V_{ij} = \delta U_{ji}^* \;. \eqno({\rm A}.9)
$$

 Making use of the elements $\delta V_{ij}$ defined through Eqs. (A.9) and 
(A.5), we can calculate from Eqs. (16) the perturbations $\delta P(\nu_\alpha 
\rightarrow \nu_\beta)$ of the neutrino--oscillation probabilities $ P(
\nu_\alpha \rightarrow \nu_\beta)$ which were evaluated before in the case of 
$ M^{(\nu)}_{12} = 0 = M^{(\nu)}_{21}$ [Eqs. (17) --- (22)].

 In particular for $\nu_e \rightarrow \nu_\mu $ oscillations, after some 
calculations up to the quadratic perturbative order in $\alpha^{(\nu)}/\mu^{(
\nu)}$, we obtain the following formula correcting Eq. (17):

%(A.10)
\begin{eqnarray*}
\delta P(\nu_e \rightarrow \nu_\mu) & \simeq & \frac{16}{841} \frac{\alpha^{(
\nu)}\alpha^{(e)}}{|m_{\nu_2}|m_\mu} \left\{ \frac{M^{(\nu)}_{33}}{m_{\nu_3}} 
\left[\sin^2(x_{21} + \varphi) + \sin^2 (x_{21} - \varphi) - 2\sin^2 \varphi
\right]\right. \\ & & \;\;\;\;\;\;\;\;\;\;\;\;\;\;\;\;\;\;\;\;\;\;\;\; - \left.
\frac{X^2}{(1 + X^2)^2}\left[\sin^2(x_{32} + \varphi) - \sin^2(x_{32} - 
\varphi)\right]\right\} \\ & & + \frac{16}{841} \left( \frac{\alpha^{(\nu)}}
{m_{\nu_2}} \right)^2 \frac{X^2}{(1 + X^2)^2} \sin^2 x_{32} \;,
\end{eqnarray*} 

\vspace{-1.53cm}

\begin{flushright}
(A.10)
\end{flushright}

\vspace{0.2cm}

\ni where $\varphi = (1/2)(\varphi^{(\nu)} - \varphi^{(e)})$. Here, $ Y = 0 $ 
for simplicity, and the upper limit of $ X \sim 0.999 $ is taken into account. 
Owing to this, in our calculations leading to Eq. (A.10), the term proportional
to $(M^{(\nu)}_{33}/m_{\nu_3})(1 - X^2|m_{\nu_2}|/m_{\nu_3}) \sim 9.04\times 
10^{-6}$ is negligible and the relation $ x_{21} \sim x_{31} \gg x_{32}$ works.
In fact, for such a value of X, we have $|m_{\nu_2}| \sim 1.05 $ eV $ \sim 
m_{\nu_3}$ ($m_{\nu_2} = -|m_{\nu_2}|$) with $m_{\nu_3}^2 - m_{\nu_2}^2 \sim 5
\times 10^{-3}\,{\rm eV}^2 $ and $|m_{\nu_2}|/m_{\nu_3} \sim 0.998 \simeq 1 $, 
while $\mu^{(\nu)} \sim 1.08\times 10^{-4}$ eV and $\beta^{(\nu)} \sim 2.20 
$~eV with $\beta^{(\nu)}/m_{\nu_3} \sim 2.09 $. Further,

%(A.11)
$$
\frac{M^{(\nu)}_{33}}{m_{\nu_3}} \sim 2.12\times 10^{-3} \sim  \frac{1 -X^2
|m_{\nu_2}|/m_{\nu_3}}{1 + X^2}\;,\; \frac{1}{1 + X^2} \sim 0.501 \;,\;
\frac{X^2}{1 + X^2} \sim 0.499
\eqno({\rm A}.11)
$$

\ni and, from Eqs. (A.4),

%(A.12)
$$
\delta m_{\nu_1} \sim -1.06 \times 10^{-5}\left( \frac{\alpha^{(\nu)}}
{m_{\nu_2}} \right)^2\;{\rm eV}\;,\; \delta m_{\nu_2,\,\nu_3}\sim \pm 0.504
\frac{\alpha^{(\nu)}}{m_{\nu_3}} \;{\rm eV}\;.  \eqno({\rm A}.12)
$$

 Since the second term in Eq. (A.10) is not invariant under the change of phase
sign, $\varphi \rightarrow -\varphi$, this term violates time reversal and so, 
CP reflection (as CPT is conserved).

 The formula (A.10) can be rewritten in the following numerical form:

%(A.13)
\begin{eqnarray*}
\delta P(\nu_e \rightarrow \nu_\mu) & \simeq & \frac{\alpha^{(
\nu)}}{|m_{\nu_2}|} \left\{4.65\times 10^{-6}\left[ \sin^2(x_{21} + \varphi) + 
\sin^2 (x_{21} - \varphi) - 2\sin^2 \varphi\right]\right. \\ 
& & \;\;\;\;\;\;\;\;\;\;\;\; -\left. 5.47\times 10^{-4}
\left[\sin^2(x_{32} + \varphi) - \sin^2(x_{32} - \varphi)\right]\right\} \\ 
& & + 4.76\times 10^{-3} \left( \frac{\alpha^{(\nu)}}{m_{\nu_2}} \right)^2 
\sin^2 x_{32} \;.
\end{eqnarray*}

\vspace{-1.53cm}

\begin{flushright}
(A.13)
\end{flushright}

\vspace{0.2cm}

\ni If $\varphi \rightarrow 0 $ and $\alpha^{(\nu)}/|m_{\nu_2}| < 1$, the sec%
ond and third term here can be neglected, when short--baseline experiments with
$ x_{21} \stackrel{<}{\sim} \pi/2 $ are taken into account, since then $ \sin^2
x_{32}\simeq x_{32}^2 \stackrel{<}{\sim} 2.5\times 10^{-5} (\pi/2)^2 $. In 
such a case, therefore, we get the two--family formula

%(A.14)
$$
\delta P(\nu_e \rightarrow \nu_\mu) \simeq 9.30\times 10^{-6} 
\frac{\alpha^{(\nu)}}{|m_{\nu_2}|} \sin^2 x_{21} \;.  \eqno({\rm A}.14)
$$

 Evidently, the oscillation amplitudes in the perturbative formula (A.13) are 
much too small to be able to help the unperturbed formula (61) in explaining 
the LSND estimate for $\nu_\mu \rightarrow \nu_e $ oscillations [9].        

\vfill\eject

~~~~
\vspace{0.6cm}

{\bf References}

\vspace{1.0cm}

{\everypar={\hangindent=0.5truecm}
\parindent=0pt\frenchspacing

{\everypar={\hangindent=0.5truecm}
\parindent=0pt\frenchspacing

~1.~W. Kr\'{o}likowski, Warsaw University preprint, December 1997 (hep--ph/%
9712328), to appear in {\it Acta Phys. Pol.}, {\bf B 29}; and references 
therein.

\vspace{0.15cm}

~2.~W.~Kr\'{o}likowski, in {\it Spinors, Twistors, Clifford Algebras and 
Quantum Deformations (Proc. 2nd Max Born Symposium 1992)}, eds. Z.~Oziewicz 
{\it et al.}, 1993, Kluwer Acad. Press; {\it Acta Phys. Pol.}, {\bf B 27}, 
2121 (1996); {\bf B 28}, 1643 (1997); Warsaw University preprint, August 1997 
(hep--ph/9709373), to appear in {\it Acta Phys. Pol.}, {\bf B 29}. 

\vspace{0.15cm}

~3.{\it ~Review of Particle Physics}, {\it Phys.Rev.} {\bf D 54}, 1 (1996), 
Part I. 

\vspace{0.15cm}

~4.~Y.~Totsuka (Super--Kamiokande Collaboration), in {\it 28th Inter. Symp. on 
Lepton--Photon Interactions}, Hamburg (Germany) 1997, to appear in the {\it 
Proceedings}.

\vspace{0.15cm}

~5.~M.~Nakahata (Super--Kamiokande Collaboration), in {\it Inter. Europhysics 
Conf. in High Energy Physics}, Jerusalem (Israel) 1997, to appear in the 
{\it Proceedings}.

\vspace{0.15cm}

~6.~M. Appolonio {\it et al.}, (CHOOZ collaboration), November 1997 
(hep--ex/9711002).

\vspace{0.15cm}

~7. S.P.~Mikheyev and A.Yu.~Smirnow, {\it Sov. Journ. Nucl. Phys.} {\bf 42}, 
913 (1986); {\it Nuovo Cimento}, {\bf C 9}, 17 (1986); L.~Wolfenstein, {\it 
Phys. Rev.} {\bf D 17}, 2369 (1978).

\vspace{0.15cm}

~8.~N.~Hata and P.~Langacker, preprint IASSNS--AST 97/29 + UPR--751 T, May 1997
(hep--ph/9705339); {\it cf.} also G.L.~Fogli, E.~Lisi and D.~Montanino, pre\-%
print~~BARI--TH/284--97 (hep--ph /9709473).

\vspace{0.15cm}

~9.~C.~Athanassopoulos {\it et al.} (LSND Collaboration), preprint UCRHEP--%
E97 (nucl--ex/9709006); and references therein.

\vfill\eject

\end{document}